\documentclass[11pt]{article}
\usepackage{amsmath}
\usepackage{amsfonts}
\usepackage{amssymb}
\usepackage{amsthm}
\usepackage{graphicx}
\usepackage{listings}
\usepackage{hyperref}
\usepackage[top=1in, bottom=1in, left=0.8in, right=0.8in]{geometry}
\usepackage{algorithm,algpseudocode}
\usepackage{tikz}              
\usetikzlibrary{arrows.meta}   
\usetikzlibrary{shapes.geometric, calc, positioning} 

\begin{document}
\title{A SAT-centered XAI method for Deep
Learning based Video Understanding}
\author{H\"ojer Key}

\maketitle

\begin{abstract}
This paper introduces a novel formal SAT-based explanation model for deep learning in video understanding. The proposed method integrates SAT solving techniques with the principles of formal explainable AI to address the limitations of existing XAI techniques in this domain. By encoding deep learning models and video data into a logical framework and formulating explanation queries as satisfiability problems, the method aims to generate logic-based explanations with formal guarantees. The paper details the conceptual framework, the process of encoding deep learning models and video data, the formulation of "Why?" and "Why not?" questions, and a novel architecture integrating a SAT solver with a deep learning video understanding model. While challenges related to computational complexity and the representational power of propositional logic remain, the proposed approach offers a promising direction for enhancing the explainability of deep learning in the complex and critical domain of video understanding.
\end{abstract}

\section{Introduction and Background}

SAT solvers have demonstrated their utility across various domains within artificial intelligence, largely due to their proficiency in managing intricate logical constraints~\cite{gong2017survey}. One notable application lies in hardware and software verification, where SAT solvers play a crucial role in validating the correctness of designs against formal specifications. For instance, in electronic design automation (EDA), SAT solvers are essential for formal equivalence checking, ensuring that different representations of a circuit design are logically consistent. They are also utilized in bounded model checking to verify that a system adheres to specific properties within a defined number of steps.

Another significant application area is automated planning and scheduling. Planning problems, such as determining a sequence of actions to achieve a particular goal, can be formulated as SAT problems. In this formulation, Boolean variables represent the state of the world at different time points, and clauses capture the preconditions and effects of actions. SAT solvers can then identify a satisfying assignment that corresponds to a valid plan. Similarly, scheduling problems, like allocating resources to tasks over time, can also be modeled and solved using SAT solvers.

SAT solvers have also found applications in other AI tasks, including constraint satisfaction problems, diagnosis, and even certain aspects of machine learning. For example, the SATzilla portfolio SAT solver employs machine learning techniques to predict the runtime of different SAT solvers on a given instance and selects the most promising solver. The success of SAT solvers in addressing computationally challenging problems across diverse AI domains underscores their versatility and potential as a fundamental tool in the field.

\subsection{Formal Explainable Artificial Intelligence (XAI): Theoretical Underpinnings}

Formal Explainable Artificial Intelligence (XAI) represents a shift in the pursuit of explainable AI~\cite{bassan2023towards}, moving towards approaches rooted in formal logic and automated reasoning, rather than relying on heuristic and often post-hoc methods. The central concept of formal XAI is to provide explanations for the behavior and decisions of AI systems, especially complex machine learning models, using rigorous logic-based definitions that can be computed and verified using automated reasoning techniques. This approach seeks to overcome the limitations of many existing XAI methods that lack formal guarantees and can sometimes produce misleading information.

\subsubsection{Logic-Based Approaches to Explainability}

Logic is fundamental in formal XAI, providing the basis for representing explanations and reasoning about the behavior of AI systems. Symbolic AI, an earlier branch of AI, heavily utilized high-level symbolic representations and logic-based reasoning. Modern formal XAI builds upon these principles, employing various logical formalisms to capture the reasoning processes of machine learning models and to formulate explanations in a precise and verifiable manner. These logical approaches offer the potential for provably correct explanations and the ability to formally reason about properties of the AI system, such as fairness or bias.

The use of logic in explainability often involves representing the classifier or other machine learning model as a logical formula. The input to the model can also be represented as a conjunction of literals, and the model's prediction as another literal. The classification process can then be viewed through the lens of logical entailment. Explanations, in this context, can take the form of logical rules or sets of conditions that are necessary or sufficient for a particular prediction.

\subsubsection{Abductive and Contrastive Explanations}

Formal XAI often differentiates between types of explanations, such as abductive and contrastive explanations. Abductive explanations aim to answer "Why?" questions by providing a set of reasons or conditions that, if true, would explain the observed outcome or prediction. For a classifier, an abductive explanation for a given instance and its predicted class might be a minimal subset of features that, when fixed to their values in the instance, guarantee the same prediction.

Contrastive explanations, conversely, address "Why not?" questions by identifying the minimal changes to the input that would lead to a different outcome. For example, a contrastive explanation might highlight which features need to be altered and by how much to change the prediction of a classifier from one class to another. Both abductive and contrastive explanations provide valuable and complementary perspectives on the model's reasoning process, allowing for a more comprehensive understanding of its behavior.

\subsubsection{Formal Verification of AI Systems}

Formal verification techniques are crucial for ensuring the correctness and reliability of AI systems, including the explanations they provide. Formal verification involves employing mathematical methods to prove that a system meets its predefined specifications or requirements. Techniques such as model checking, automated theorem proving, and deductive verification can be applied to analyze AI algorithms and their implementations to ensure they behave as expected under all possible conditions.

In the context of explainability, formal verification can be used to assess the soundness and completeness of the generated explanations. Soundness would imply that if the explanation claims a certain set of factors led to a prediction, then those factors indeed guarantee that prediction according to the model's logic. Completeness would mean that if a valid explanation exists within the defined framework, the verification process would be able to find it. While formal verification offers a high degree of rigor, it can be challenging to apply to very complex AI systems, like deep neural networks, due to their intricate nature and vast state spaces. Nevertheless, it remains a crucial aspect of ensuring the trustworthiness and reliability of both AI systems and their explanations.

\section{Related Work}

\subsection{Combining SAT Solving and Formal XAI for Machine Learning Explanation}

The integration of SAT solving techniques with the principles of formal XAI has emerged as a promising direction for generating rigorous explanations for machine learning models. By leveraging the power of SAT solvers as automated reasoners, researchers have explored methods to extract logical explanations from various types of machine learning models, including classifiers.

\subsubsection{Using SAT for Explaining Classifiers and Other ML Models}

SAT solvers can be employed to explain the predictions of classifiers by formulating the classification process and the explanation query as a satisfiability problem. For instance, to find an abductive explanation for a specific prediction, one could encode the classifier's behavior as a set of logical clauses. Then, the input instance and the predicted class are also encoded. The explanation query could be formulated as finding a minimal subset of the input features such that, when their values are fixed, the resulting SAT problem (representing the classifier and the fixed features) still entails the predicted class. If a satisfying assignment exists under these conditions, it indicates that the chosen subset of features is sufficient to guarantee the prediction. SAT solvers can then be used to find such minimal subsets, providing an explanation in terms of the most relevant input features.

Similarly, contrastive explanations can be sought by encoding the condition that the model should predict a different class. The SAT solver can then identify the minimal changes to the input features that satisfy this condition, thus providing a "why not?" explanation. This approach has been applied to various interpretable machine learning models, such as decision trees and rule-based systems, as well as to more complex models like neural networks through appropriate encodings.

\subsubsection{Logic-Based Rule Extraction from Machine Learning Models}

Another way SAT solvers can contribute to explainability is through logic-based rule extraction from trained machine learning models. The goal of rule extraction is to identify a set of human-readable logical rules that accurately capture the decision-making logic of the model. SAT solvers can be used in this process to find minimal sets of rules that are consistent with the model's predictions on the training data. For example, the model's behavior on the training set can be encoded as a SAT problem, where variables represent features and their values, and clauses represent the model's predictions. By finding satisfying assignments to this SAT problem under certain constraints (e.g., minimizing the number or complexity of the rules), one can extract a set of logical rules that effectively mimic the model's behavior and can serve as explanations for its predictions.

These extracted rules can provide a more transparent and understandable representation of the model's internal logic compared to the complex mathematical operations within the model itself. They can highlight the key features and their combinations that lead to specific predictions, making it easier for humans to understand and trust the model's decisions.

\subsection{Explainable AI for Deep Learning Models}

Explainable AI for deep learning models is a rapidly evolving field that explores various techniques to shed light on the decision-making processes of these complex architectures. While many of these techniques are not formal in nature, understanding the current landscape is crucial for positioning the proposed SAT-based approach.

\subsubsection{Post-hoc Explanation Techniques for Neural Networks}

A significant portion of research in deep learning explainability focuses on post-hoc methods, which are applied to a trained model to understand its behavior without necessarily altering its architecture or training process. Popular post-hoc techniques include:

\begin{itemize}
    \item \textbf{Saliency Maps:} These methods compute a relevance score for each input feature (e.g., pixels in an image, words in a text) indicating its importance for the model's prediction. Various techniques exist, such as gradient-based methods, which highlight the input regions that have the strongest influence on the output.
    \item \textbf{Attention Mechanisms:} Many modern deep learning models, particularly in areas like natural language processing and video understanding, utilize attention mechanisms. These mechanisms learn to weigh the importance of different parts of the input when making a prediction. The attention weights can often be interpreted as indicating which input elements the model focused on, providing a form of explanation.
    \item \textbf{LIME (Local Interpretable Model-agnostic Explanations):} LIME works by approximating the behavior of the complex deep learning model locally around a specific prediction using a simpler, interpretable model (e.g., a linear model or a decision tree). By analyzing this local surrogate model, one can gain insights into which input features were most important for that particular prediction.
    \item \textbf{SHAP (SHapley Additive exPlanations):} SHAP is a method based on game theory that aims to quantify the contribution of each input feature to the model's output for a given instance~\cite{sundararajan2020many}. It computes Shapley values, which represent the average marginal contribution of a feature across all possible feature combinations.
\end{itemize}

These techniques provide valuable insights into the behavior of deep learning models, often offering intuitive visual or feature-based explanations. However, they often lack formal guarantees and can be sensitive to the specific implementation or parameters used.

\subsubsection{Limitations of Existing Methods in Providing Formal Guarantees}

A major limitation of many existing XAI methods for deep learning is the absence of formal guarantees regarding the correctness or completeness of the explanations they provide. While these methods can often highlight potentially relevant features or regions, they do not necessarily provide a logical justification for the model's prediction in a way that can be formally verified. This lack of rigor can be problematic, especially in critical applications where trust and accountability are paramount. For instance, a saliency map might highlight a certain region in an image, but it does not formally prove that the model's decision was solely based on that region or explain the underlying reasoning process. The potential for misleading information from informal XAI methods has been demonstrated, underscoring the need for more rigorous approaches.

\subsection{Explainable AI in the Domain of Video Understanding}

Explainable AI in the domain of video understanding presents unique challenges due to the spatio-temporal nature of video data and the complexity of deep learning models used for video analysis.

\subsubsection{Current Techniques and Their Shortcomings}

Extending XAI techniques to video understanding often involves adapting existing methods to handle the temporal dimension or developing new techniques specifically designed for video data. Some common approaches include:

\begin{itemize}
    \item \textbf{Spatio-Temporal Attention:} Models that utilize attention mechanisms can provide insights into both the spatial regions and the temporal segments that the model deems important for its prediction. Visualizing these attention weights across frames can offer a spatio-temporal explanation.
    \item \textbf{Gradient-Based Methods for Video:} Gradient-based saliency methods can be extended to video by computing gradients with respect to both spatial and temporal inputs. This can highlight the video frames or regions within frames that are most influential for the prediction.
    \item \textbf{Temporal Perturbation Analysis:} This involves perturbing or masking different temporal segments of the video and observing the impact on the model's prediction. By identifying which segments are most critical, one can gain insights into the model's temporal reasoning.
\end{itemize}

While these techniques offer some level of explanation for video understanding models, they often face limitations in providing comprehensive and formal explanations for video-level predictions. Explaining how long-term temporal relationships between events contribute to the final prediction remains a significant challenge. Furthermore, the high computational cost of analyzing and explaining complex video models can be a limiting factor.

\subsubsection{Challenges Specific to Explaining Spatio-Temporal Models}

Explaining models that process both spatial and temporal information in videos introduces several specific challenges:

\begin{itemize}
    \item \textbf{Attribution to Specific Regions and Segments:} It is difficult to precisely attribute a prediction to specific spatial regions within individual frames and to specific temporal segments across the entire video. The model's decision might be based on a complex interplay of both spatial and temporal features.
    \item \textbf{Explaining Temporal Relationships:} Understanding how the temporal relationships between different events or movements in a video contribute to the final prediction is particularly challenging. Many existing methods struggle to capture these complex temporal dependencies in their explanations.
    \item \textbf{Computational Complexity:} Analyzing and explaining deep learning models for video, which often involve processing large amounts of high-dimensional data, can be computationally very expensive. This limits the scalability and applicability of some explanation techniques.
\end{itemize}

\subsubsection{State-of-the-Art Deep Learning Models for Video Understanding and Their Explainability Bottlenecks}

State-of-the-art deep learning models for video understanding often employ complex architectures such as 3D Convolutional Neural Networks (CNNs), recurrent neural networks (RNNs) like LSTMs and GRUs, and increasingly, Transformer-based models. These models can capture intricate spatio-temporal patterns but are often deeply layered and highly non-linear, making their internal representations and decision-making processes very difficult to interpret.

For instance, 3D CNNs extend 2D convolutions to the temporal dimension, allowing them to learn spatio-temporal features directly from video data. However, understanding which specific spatio-temporal features are most important for a given prediction can be challenging. RNNs can model temporal dependencies, but the information is processed sequentially through hidden states, making it difficult to trace back the influence of specific past events on the current prediction. Transformer-based models, while powerful in capturing long-range dependencies, often have a large number of parameters and complex attention mechanisms that can be hard to interpret.

The inherent complexity of these state-of-the-art models presents significant explainability bottlenecks. Their distributed representations and non-linear interactions make it hard to understand the causal relationships between the input video and the final prediction. This lack of transparency hinders trust, debugging, and further improvement of these powerful video understanding systems.

\section{Proposed Method: Formal SAT-Based Explanation Model for Deep Video Understanding}

This section introduces a novel method for providing formal, logic-based explanations for deep learning models applied to video understanding tasks. The proposed approach integrates SAT solving techniques with the principles of formal XAI to address the limitations of existing explanation methods in this domain.

\subsection{Conceptual Framework: Integrating SAT Solving and Formal XAI for Video Analysis}

The core idea of the proposed method is to leverage the rigorous reasoning capabilities of SAT solvers to explain the predictions made by deep learning models on video data. This is achieved by creating a formal, logical representation of the deep learning model and the input video, and then formulating explanation queries as satisfiability problems within this logical framework. By solving these SAT problems, we can extract logical explanations for the model's behavior.

The overall conceptual framework involves the following key steps:

\begin{enumerate}
    \item  \textbf{Formal Encoding of the Deep Learning Model:} A chosen deep learning model for video understanding is translated into a set of Boolean variables and logical clauses. This encoding captures the model's architecture, parameters (weights and biases), and activation functions.
    \item  \textbf{Formal Representation of Video Data:} The input video is represented within the same logical framework. This might involve encoding features extracted from the video frames and the temporal relationships between them.
    \item  \textbf{Formulation of Explanation Queries as SAT Problems:} Specific explanation queries are translated into logical constraints that are added to the encoded model and video data. For example, a "Why?" query might involve finding a minimal set of input features or internal activations that are necessary for the observed prediction. A "Why not?" query could involve finding minimal changes to the input that would lead to a different prediction.
    \item  \textbf{SAT Solving and Explanation Extraction:} A SAT solver is then used to find a satisfying assignment to the resulting set of logical clauses. The solution provided by the SAT solver is interpreted back in the context of the original model and video data to yield a formal explanation.
\end{enumerate}

This framework aims to provide explanations that are not only interpretable but also formally verifiable, addressing the lack of rigor in many existing XAI techniques for deep learning in video understanding.

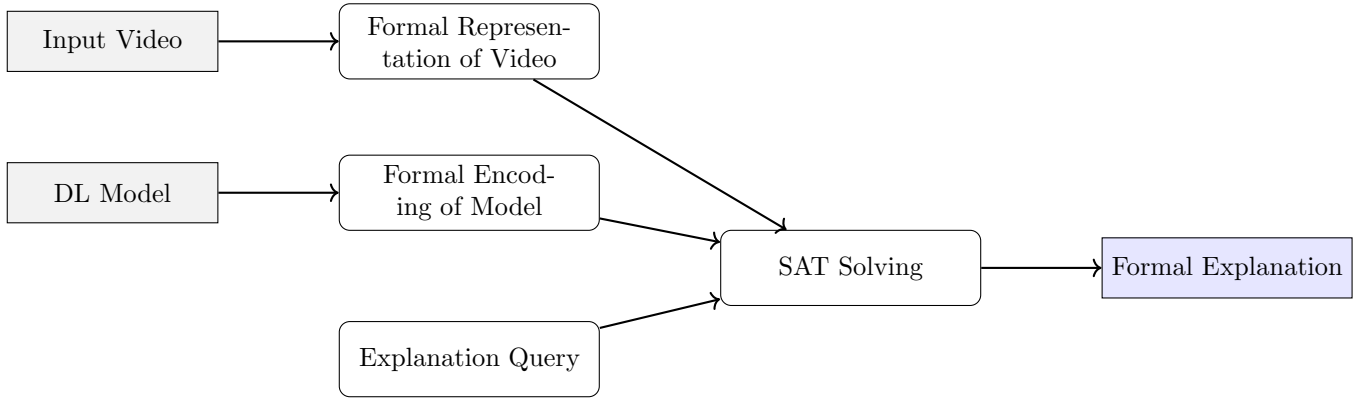
\begin{figure}[h!]
    \centering
\begin{tikzpicture}[
    node distance=1.2cm and 1.6cm,
    every node/.style={align=center},
    process/.style={rectangle, draw, rounded corners, minimum height=1.0cm, minimum width=3.0cm, text width=3.2cm, font=\small},
    input/.style={rectangle, draw, fill=gray!10, minimum height=0.8cm, minimum width=2.8cm, font=\small},
    output/.style={rectangle, draw, fill=blue!10, minimum height=0.8cm, minimum width=3.0cm, font=\small},
    arrow/.style={->, thick}
]

\node[input] (video) {Input Video};
\node[process, right=of video] (videoEncoding) {Formal Representation of Video};

\node[input, below=of video] (dlmodel) {DL Model};
\node[process, right=of dlmodel] (modelEncoding) {Formal Encoding of Model};

\node[process, below=of modelEncoding] (queryFormulation) {Explanation Query};

\node[process, right=of modelEncoding, yshift=-1cm] (satSolving) {SAT Solving};

\node[output, right=of satSolving] (explanation) {Formal Explanation};

\draw[arrow] (video) -- (videoEncoding);
\draw[arrow] (dlmodel) -- (modelEncoding);
\draw[arrow] (videoEncoding) -- (satSolving);
\draw[arrow] (modelEncoding) -- (satSolving);
\draw[arrow] (queryFormulation) -- (satSolving);
\draw[arrow] (satSolving) -- (explanation);

\end{tikzpicture}
    \caption{High-Level Architecture of the Proposed SAT-Based Explanation Model for Deep Video Understanding}
    \label{fig:architecture}
\end{figure}

Figure \ref{fig:architecture} illustrates the high-level architecture of the proposed system. It comprises a pre-trained deep learning model for video understanding, an encoding module that translates the model and the input video into a SAT problem, a query formulation module that defines the explanation query as logical constraints, a SAT solver, and an explanation extraction module that interprets the solver's output.

\subsection{Formal Representation of Deep Learning Models for Video}

The first crucial step in the proposed method is to develop a formal way to represent the deep learning model used for video understanding within a logical framework. This involves encoding both the architecture and the parameters of the neural network using propositional logic.

\subsubsection{Encoding Neural Network Architectures and Parameters in a Logical Framework}

One possible approach to encoding a deep learning model in propositional logic involves discretizing the continuous activations and weights of the network. For a given pre-trained deep learning model, each neuron's activation and each weight connecting neurons can be represented by a set of Boolean variables. The number of Boolean variables used to represent each continuous value would determine the precision of the discretization.

For example, a real-valued activation $a$ could be represented by $n$ Boolean variables $b_1, b_2, \dots, b_n$, where the binary sequence $b_n b_{n-1} \dots b_1$ represents a quantized version of $a$ within a certain range and precision. Similarly, weights and biases can also be discretized and represented using Boolean variables.

The operations performed by the neurons, such as weighted sums and activation functions, need to be translated into logical clauses that operate on these Boolean variables. For a neuron $j$ in layer $l$, its activation $a_j^{(l)}$ is typically computed as
\[
a_j^{(l)} = f\left( \sum_i w_{ij}^{(l)} a_i^{(l-1)} + b_j^{(l)} \right),
\]
where $f$ is the activation function, $w_{ij}^{(l)}$ is the weight connecting neuron $i$ in layer $l-1$ to neuron $j$ in layer $l$, and $b_j^{(l)}$ is the bias.

To encode this in propositional logic, the discretized representations of the input activations $a_i^{(l-1)}$ and the weight $w_{ij}^{(l)}$ would be used to compute the discretized representation of the weighted sum $\sum_i w_{ij}^{(l)} a_i^{(l-1)} + b_j^{(l)}$. This summation and multiplication of discretized values can be implemented using Boolean logic gates (AND, OR, NOT), which can then be expressed as Conjunctive Normal Form (CNF) clauses using standard transformations. The activation function $f$ also needs to be approximated using Boolean logic based on the discretized input. For common activation functions like ReLU, sigmoid, or tanh, piecewise linear approximations can be used, and each linear segment can be encoded using Boolean variables and clauses.

Consider a simplified example: a neuron with two binary inputs $x_1, x_2$, weights $w_1 = 1, w_2 = 1$, bias $b = 0$, and a threshold activation function that outputs 1 if the weighted sum is $\geq 2$ and 0 otherwise. The weighted sum is $x_1 + x_2$. The output is 1 only if $x_1 = 1$ and $x_2 = 1$. This can be directly represented by the clause $(x_1 \wedge x_2) \Leftrightarrow y$, which in CNF is $(\neg x_1 \vee \neg x_2 \vee y) \wedge (x_1 \vee \neg y) \wedge (x_2 \vee \neg y)$.

For more complex scenarios with discretized real values, the encoding would involve a larger number of Boolean variables and clauses to represent the arithmetic operations and the activation functions with sufficient precision. The level of abstraction for the encoding can also be adjusted. For instance, instead of encoding individual neurons, one could potentially encode the overall input-output behavior of certain layers or modules using logical rules learned from the trained network.

\subsubsection{Handling Spatio-Temporal Dependencies in Video Data}

Video data introduces the challenge of representing both spatial information within each frame and the temporal relationships between consecutive frames. In the proposed method, the input video would first be pre-processed to extract relevant features. These features could be high-level semantic features obtained from pre-trained models or lower-level features like motion vectors or visual descriptors.

Each extracted feature at each time step (frame) would then be represented within the logical framework, possibly using a similar discretization approach as used for the neural network's activations and weights. For example, if the video understanding model operates on a sequence of feature vectors extracted from the frames, each element of these feature vectors at each time step would be represented by a set of Boolean variables.

The temporal connections between frames could be handled by explicitly representing the flow of information through the deep learning model over time. For models like RNNs or LSTMs, the state transitions between time steps would need to be encoded as logical clauses. For models like 3D CNNs or Transformer networks that process the entire video or chunks of it at once, the spatial and temporal relationships are implicitly learned within the network's architecture. The encoding would then need to capture these learned relationships in the logical framework.

For example, if the video understanding model uses a 3D convolutional layer that operates on a spatio-temporal neighborhood of voxels, the logical encoding would need to represent the input features within this neighborhood and the convolutional operation that combines them to produce the output features. This would involve encoding the weights of the convolutional kernel and the summation and non-linear activation.

\subsection{SAT-Based Query Formulation for Explainability}

Once the deep learning model and the video data are represented within the logical framework, the next step is to formulate the explanation queries as satisfiability problems. This involves adding logical constraints to the encoded system that correspond to the specific question being asked.

\subsubsection{Defining Explanation Queries as Satisfiability Problems}

Both "Why?" (abductive) and "Why not?" (contrastive) questions can be formulated as SAT problems within this framework.

For a "Why?" question, such as "Why was this action recognized as 'jumping' in this video?", we are looking for a minimal set of conditions (in terms of input features or internal activations) that are necessary for the model to predict 'jumping'. This can be formulated as a SAT problem by:

\begin{enumerate}
    \item  Encoding the deep learning model and the input video.
    \item  Adding a constraint that the output of the model is 'jumping'.
    \item  Iteratively trying to find a minimal subset of the input features or internal activations such that, if these are fixed to their values in the original video, the SAT solver still finds a satisfying assignment (meaning the prediction of 'jumping' holds).
    \item  The minimal set of fixed features or activations would then constitute the explanation.
\end{enumerate}

For a "Why not?" question, such as "Why was the action not recognized as 'running' instead of 'jumping'?", we want to find the minimal changes to the input video that would cause the model to predict 'running' instead of 'jumping'. This can be formulated as a SAT problem by:

\begin{enumerate}
    \item  Encoding the deep learning model and the input video.
    \item  Adding a constraint that the output of the model is 'running'.
    \item  Adding a constraint that the input video is as similar as possible to the original video (e.g., by minimizing the number of feature values that are changed).
    \item  The SAT solver would then find an assignment to the input features that leads to the prediction of 'running' while being minimally different from the original input. The differences would constitute the contrastive explanation.
\end{enumerate}

\subsubsection{Formulating "Why?" and "Why Not?" Questions using Logical Constraints}

The explanation queries are formalized by adding specific logical constraints to the SAT problem. These constraints depend on the type of explanation being sought.

For a "Why?" explanation focusing on input features, the constraints would involve fixing a subset of the Boolean variables representing the input video features to their observed values and checking if the SAT problem (model + fixed features + target prediction) is still satisfiable. Minimality can be achieved by iteratively reducing the size of this subset.

Mathematically, let $M$ be the set of clauses representing the encoded deep learning model, and let $V$ be the set of clauses representing the encoded input video features. Suppose the model predicts class $c$ for this video. To find a minimal subset of input features $F \subseteq V$ that are necessary for this prediction, we can formulate the SAT problem
\[
M \wedge F \wedge (\text{output} = c).
\]
We then aim to find a minimal $F$ for which this problem is satisfiable.

For a "Why not?" explanation, suppose the model predicted class $c_1$ for input $V$, and we want to know why it didn't predict class $c_2$. We can formulate a SAT problem
\[
M \wedge V' \wedge (\text{output} = c_2),
\]
where $V'$ represents a modified version of the input video. The goal is to find a $V'$ that satisfies this formula and minimizes the difference between $V$ and $V'$ according to some distance metric. This difference would represent the minimal change needed to flip the prediction to $c_2$. The distance metric can be incorporated into the SAT problem by introducing additional variables and clauses that penalize changes to the original input features.

For example, if a feature is represented by $n$ Boolean variables, changing its value could be represented by flipping one or more of these bits. We can add constraints that minimize the number of flipped bits while still satisfying the condition of predicting class $c_2$.

\subsection{Novel Model Architecture: Integrating a SAT Solver with a Deep Learning Video Understanding Model}

The proposed model architecture integrates a pre-trained deep learning model for video understanding with a SAT solver to generate formal explanations.

\subsubsection{Detailed Description of the Architecture and its Components}

The architecture consists of the following main components:

\begin{enumerate}
    \item  \textbf{Pre-trained Deep Learning Model for Video Understanding:} This is the model whose predictions we want to explain. It could be any state-of-the-art model suitable for the video understanding task at hand (e.g., a 3D CNN for action recognition).
    \item  \textbf{Encoding Module:} This module takes the pre-trained deep learning model and an input video as input and translates them into a SAT problem represented as a CNF formula. This involves discretizing the model's weights, biases, and activations, as well as the features extracted from the video, and representing the model's operations and the video data using Boolean variables and logical clauses.
    \item  \textbf{Query Formulation Module:} This module takes an explanation query (e.g., "Why this prediction?", "Why not that prediction?") and formulates it as a set of logical constraints that are added to the SAT problem generated by the encoding module.
    \item  \textbf{SAT Solver:} A high-performance SAT solver (e.g., a CDCL solver) is used to find a satisfying assignment to the complete SAT problem (encoded model + encoded video + query constraints).
    \item  \textbf{Explanation Extraction Module:} This module takes the output of the SAT solver (a satisfying assignment or an indication of unsatisfiability) and interprets it back in the context of the original deep learning model and video data to generate a formal explanation. The explanation could be in the form of a minimal set of necessary input features, a minimal change to the input that would flip the prediction, or a set of logical rules that justify the prediction.
\end{enumerate}

The flow of information in this architecture is as follows: First, an input video is fed into the encoding module along with the pre-trained deep learning model. The encoding module generates a CNF formula representing the model and the video. Then, an explanation query is provided to the query formulation module, which adds logical constraints to the CNF formula based on the query. This complete SAT problem is then passed to the SAT solver. The SAT solver attempts to find a satisfying assignment. If a satisfying assignment is found, the explanation extraction module interprets this assignment to produce the explanation. If the SAT solver determines that the problem is unsatisfiable, it might indicate that no explanation of the requested type exists within the defined framework.

\subsubsection{Workflow of the Explanation Generation Process}

The step-by-step workflow of the explanation generation process is as follows:

\begin{enumerate}
    \item  \textbf{Input:} A pre-trained deep learning model for video understanding, an input video, and an explanation query.
    \item  \textbf{Encoding:} The encoding module discretizes the weights, biases, and activations of the deep learning model. It also extracts features from the input video and discretizes them. Both the model's operations and the video features are then translated into a CNF formula using Boolean variables and logical clauses.
    \item  \textbf{Query Formulation:} Based on the explanation query (e.g., "Why was the action classified as 'running'?"), the query formulation module adds logical constraints to the CNF formula. For a "Why?" query, this might involve constraining the output variables to represent the predicted class. For a "Why not?" query, it might involve constraining the output variables to represent an alternative class and adding constraints to minimize the difference between the original and a modified input.
    \item  \textbf{SAT Solving:} The complete CNF formula (encoded model + encoded video + query constraints) is given as input to a SAT solver. The SAT solver searches for a satisfying assignment of truth values to the Boolean variables.
    \item  \textbf{Explanation Extraction:}
    \begin{itemize}
        \item   If the SAT solver finds a satisfying assignment for a "Why?" query aimed at finding necessary input features, the explanation extraction module identifies the minimal set of input feature variables that must be set to their original values in order to satisfy the formula. These features are then presented as the explanation.
        \item   If the SAT solver finds a satisfying assignment for a "Why not?" query, the explanation extraction module compares the values of the input feature variables in the satisfying assignment to their original values. The differences represent the minimal changes to the input that would lead to the alternative prediction, and these differences are presented as the explanation.
        \item   If the SAT solver determines that the SAT problem is unsatisfiable, it might indicate that no explanation of the requested type exists within the defined encoding and query constraints.
    \end{itemize}
    \item  \textbf{Output:} A formal explanation for the deep learning model's prediction on the input video, based on the SAT solver's result.
\end{enumerate}

\section{Technical Details and Proof}

This section delves into the technical and mathematical aspects of the proposed method, providing formal definitions, details of the logical encodings, and algorithms for explanation generation.

\subsection{Technical and Mathematical Details}

This section delves into the technical and mathematical aspects of the proposed method, providing formal definitions, details of the logical encodings, and algorithms for explanation generation.

\subsubsection{Formal Definitions of Key Concepts}

\begin{itemize}
    \item   \textbf{Discretized Neural Network State:} The state of a deep learning model (weights, biases, activations) is represented by a set of Boolean variables obtained through a discretization process. Let $W$ be the set of weights, $B$ the set of biases, and $A$ the set of activations in the original model. Their discretized counterparts are represented by sets of Boolean variables $B_W$, $B_B$, $B_A$.

    \item   \textbf{Discretized Video Features:} Features extracted from the input video at different time steps are also discretized and represented by a set of Boolean variables $B_F$.

    \item   \textbf{Logical Encoding Function:} A function $E$ that maps the deep learning model and the input video to a CNF formula $\phi = E(\text{Model}, \text{Video})$ over the Boolean variables $B_W \cup B_B \cup B_A \cup B_F \cup B_O$, where $B_O$ represents the discretized output of the model.

    \item   \textbf{Explanation Query Constraints:} A set of logical clauses $Q$ that represent the specific explanation query (e.g., target prediction, constraints on input features).

    \item   \textbf{SAT Problem for Explanation:} The satisfiability problem is defined by the CNF formula $\phi \wedge Q$. A satisfying assignment to this formula provides the basis for extracting the explanation.
\end{itemize}

\subsubsection{Logical Encodings and Transformations}

The logical encoding function $E$ needs to capture the behavior of each layer and operation in the deep learning model. For a linear layer with weights $w_{ij}$ and inputs $x_i$, the output $y_j = \sum_i w_{ij} x_i$ is first computed using discretized representations of $w_{ij}$ and $x_i$. The multiplication and addition of discretized values can be implemented using Boolean logic.

For example, consider the multiplication of two discretized values. If $x$ and $w$ are each represented by $n$ bits, the multiplication can be implemented using a series of bitwise AND and shift operations, similar to how multiplication is done in digital circuits. Each bit of the product can be represented by a Boolean variable, and the relationships between these variables and the input bits can be expressed as CNF clauses. Addition can be implemented using full adders, where each bit of the sum and the carry-out bit are represented by Boolean variables, and their relationships with the input bits and carry-in bit are encoded as CNF clauses.

The activation function applied to the output of the linear layer also needs to be encoded. For a ReLU function ($\max(0, z)$), if $z$ is represented by $n$ Boolean variables, the output will be 0 if $z \leq 0$ and $z$ if $z > 0$. This can be encoded by checking the sign bit of the discretized $z$. If the sign bit indicates negative, the output Boolean variables are all set to false (representing 0). Otherwise, they are set to the same values as the Boolean variables representing $z$. Similar encodings can be developed for other activation functions using piecewise linear approximations or by directly implementing their logical equivalents on the discretized representations.

For video data, if features are extracted as real-valued vectors at each time step, each element of these vectors is discretized into a set of Boolean variables. The temporal relationships, if modeled explicitly (e.g., in an RNN), would require encoding the state transitions between time steps. This would involve representing the hidden state at each time step using Boolean variables and encoding the update rules of the RNN (e.g., for LSTM, the update of the cell state and hidden state) using logical clauses based on the discretized inputs, previous hidden state, and weights.

\subsubsection{Algorithms for Explanation Generation using SAT Solvers}

\begin{algorithm}
\caption{Finding Minimal Necessary Input Features ("Why?" Explanation)}
\label{alg:why_explanation}
\begin{algorithmic}[1]
\Require Deep learning model, input video
\Ensure Minimal set of necessary input features $E$
\State Encode the deep learning model and the input video into a CNF formula $\phi$ 
\State Add a constraint $C$ that the output of the model is the predicted class $c$. Let $\phi' = \phi \wedge C$ 
\State Let $F$ be the set of Boolean variables representing the input video features
\State Initialize an explanation set $E = F$
\For{each feature variable $f \in F$}
    \State Let $F' = E \setminus \{f\}$
    \State Form a new SAT problem $\phi''$ by adding constraints to $\phi'$ that fix all variables in $F'$ to their values in the original encoded video 
    \State Run a SAT solver on $\phi''$
    \If{$\phi''$ is satisfiable}
        \State Update $E = F'$
    \EndIf
\EndFor
\State \textbf{return} $E$
\end{algorithmic}
\end{algorithm}

\begin{algorithm}
\caption{Finding Minimal Changes for an Alternative Prediction ("Why Not?" Explanation)}
\label{alg:why_not_explanation}
\begin{algorithmic}[1]
\Require Deep learning model, input video
\Ensure Minimal changes to input features
\State Encode the deep learning model and the input video into a CNF formula $\phi$ 
\State Let $c_1$ be the predicted class for the original video
\State Let $c_2$ be the alternative class we want to understand why it wasn't predicted 
\State Add a constraint $C'$ that the output of the model is $c_2$. Let $\phi''' = \phi \wedge C'$ 
\State Introduce additional Boolean variables to represent potential flips in the bits of the discretized input features
\State Add clauses to $\phi'''$ that relate the original input feature variables to their flipped versions and to the output $c_2$
\State Add an objective function to the SAT solver that minimizes the number of flipped bits (using techniques like iterative deepening or MaxSAT if the solver supports it)
\State Run the SAT solver with the objective function
\If{a satisfying assignment is found}
    \State \textbf{return} Flipped bits in the input features
\EndIf
\end{algorithmic}
\end{algorithm}

\subsubsection{Mathematical Proofs}

Proving the soundness and completeness of this approach depends heavily on the accuracy of the logical encoding of the deep learning model and the video data.

\paragraph{Soundness (Sketch):}

If Algorithm \ref{alg:why_explanation} finds a set of necessary input features $E$, and the SAT solver returns a satisfying assignment when only these features are fixed (along with the model and the target prediction), then it implies that the model's internal logic, as captured by the encoding, can still lead to the target prediction based solely on these features. The encoding aims to be a faithful representation of the discretized model's behavior.

\paragraph{Completeness (Sketch):}

Completeness would imply that if there exists a minimal set of necessary features (or minimal changes for a contrastive explanation), Algorithm \ref{alg:why_explanation} (or \ref{alg:why_not_explanation}) will find it. This depends on the SAT solver's ability to find a satisfying assignment if one exists and on the effectiveness of the iterative minimization process in Algorithm \ref{alg:why_explanation} or the optimization process in Algorithm \ref{alg:why_not_explanation}. For Algorithm \ref{alg:why_explanation}, the iterative removal of features ensures that the resulting set is minimal with respect to the features considered. For Algorithm \ref{alg:why_not_explanation}, if the SAT solver with the optimization objective finds a solution, it will be one with the minimum number of changes within the limitations of the encoding and the solver's capabilities.

Formal proofs would require a more rigorous definition of the encoding function $E$ and the relationship between the continuous model and its discrete logical representation. This could involve bounding the error introduced by discretization and showing that the logical clauses correctly implement the intended arithmetic and logical operations.

\section{Discussions}

This section discusses the implications of the proposed formal SAT-based explanation model for deep video understanding, its potential advantages and limitations, and outlines promising directions for future research.

\subsection{In-depth Discussion of the Proposed Method and its Implications}

The proposed method offers a novel approach to explainability in deep learning for video understanding by leveraging the rigor of formal methods and the power of SAT solving.

\subsubsection{Advantages over Existing XAI Techniques for Video Understanding}

Compared to many existing XAI techniques for video understanding, such as spatio-temporal attention or gradient-based saliency maps, the proposed method has the potential to provide explanations with formal guarantees. While visual explanations can be intuitive, they often lack a clear logical justification for the model's prediction. The SAT-based approach, by encoding the model's behavior and the explanation query in a logical framework, can provide explanations that are rooted in logical deduction. For "Why?" explanations, the identified necessary input features are those that, according to the encoded model's logic, are essential for achieving the prediction. For "Why not?" explanations, the minimal changes found are those that, when applied to the input, satisfy the logical constraints of the model predicting the alternative outcome. This contrasts with heuristic methods that might highlight features that are merely correlated with the prediction rather than being logically necessary or sufficient.

Furthermore, the proposed method can potentially address the limitations of current techniques in explaining long-term temporal reasoning in videos. By encoding the temporal dynamics of the deep learning model (e.g., through the encoding of recurrent layers or by considering the entire spatio-temporal input for models like 3D CNNs), the SAT solver can reason about how events across different time steps contribute to the final prediction. This could lead to more comprehensive explanations for video-level tasks that involve understanding sequences of actions or events over time.

\subsubsection{Addressing the Limitations of Current Deep Learning Models in Terms of Explainability}

The inherent complexity of deep learning models often makes it difficult to understand their decision-making processes. The proposed method attempts to overcome this by providing a different lens through which to analyze these models – the lens of formal logic. By successfully encoding a deep learning model into a SAT problem, we are essentially creating a symbolic representation of its behavior. The explanations generated by querying this symbolic representation using a SAT solver can offer insights into the model's logic that might be hard to obtain by directly analyzing the model's weights and activations.

For instance, the minimal set of necessary input features identified for a prediction can highlight which aspects of the video the model deemed crucial. The minimal changes required to flip a prediction can reveal the model's sensitivity to certain features and the decision boundaries it has learned. This level of insight can be valuable for understanding the model's strengths and weaknesses, identifying potential biases, and guiding efforts to improve its performance or robustness.

\subsubsection{Potential Applications and Impact of the Proposed Model}

The proposed explainable video understanding model has the potential to be applied in various domains where understanding the reasoning behind a model's predictions is critical. In surveillance, for example, it could explain why a particular activity was flagged as suspicious, potentially highlighting the specific movements or interactions that led to the alert. In autonomous driving, it could provide explanations for why the system made a certain driving decision in a complex scenario, which could be crucial for safety and accountability. In medical diagnosis from video data (e.g., analyzing patient movements or surgical procedures), the model could explain the reasoning behind a diagnosis, aiding medical professionals in understanding and trusting the AI's assessment. In content analysis, it could explain why a video was classified in a certain way, providing insights into the features the model considers important for different categories.

The ability to provide formal explanations can significantly impact the trust and accountability associated with using deep learning in these areas. When users can understand the logical reasons behind a model's predictions, they are more likely to trust its decisions . Furthermore, if errors occur, the explanations can help in debugging the model and identifying the root causes of the failures, leading to more effective improvements.

\subsubsection{Challenges and Limitations of the Proposed Approach}

Despite its potential advantages, the proposed method also faces several challenges and limitations that need to be considered.

\paragraph{Computational Complexity and Scalability Considerations}

Encoding a deep learning model, especially a large and complex one used for video understanding, into a SAT problem can be computationally very expensive. The number of Boolean variables and clauses required to represent the model's weights, biases, activations, and operations can grow rapidly with the size and complexity of the network. Similarly, encoding a video, even after feature extraction, can also lead to a large number of variables and clauses, especially when considering temporal dependencies.

Solving the resulting SAT problem can also be computationally intensive, as SAT is an NP-complete problem. While modern SAT solvers have made significant progress and can handle instances with millions of variables and clauses, the size and complexity of the SAT problems arising from encoding deep learning models for video might still pose a significant scalability challenge. Future work needs to explore efficient encoding techniques and investigate the use of optimized SAT solvers, potentially including parallel solvers, to address these issues. Abstractions or approximations in the encoding process might also be necessary to achieve practical scalability.

\paragraph{Representational Power of the Chosen Logical Framework}

The proposed method primarily relies on propositional logic for encoding the deep learning model and video data. While propositional logic is powerful, it might have limitations in fully capturing the continuous and nuanced reasoning of deep learning models. The discretization process, while necessary to translate continuous values into the Boolean domain, inherently introduces some level of approximation. The precision of this discretization will affect the accuracy of the logical representation. A higher precision would require more Boolean variables, leading to a larger SAT problem, while a lower precision might not accurately reflect the model's behavior.

Exploring the use of more expressive logical frameworks, such as Satisfiability Modulo Theories (SMT), which can handle theories like arithmetic and inequalities, might offer a way to represent the continuous nature of deep learning models more directly. However, SMT solving can also be computationally challenging. Future research could investigate the trade-offs between the expressiveness of the logical framework and the scalability of the explanation generation process.

\paragraph{Potential Trade-offs between Explainability and Model Accuracy}

There is a potential for a trade-off between making a deep learning model more explainable through the proposed integration and its accuracy or performance on the video understanding task. The discretization of the model and the video data might lead to some loss of information, potentially affecting the accuracy of the encoded model compared to the original continuous model. Furthermore, any approximations made in encoding the activation functions or other non-linear operations could also impact the fidelity of the logical representation.

It is important to carefully consider the level of discretization and the types of approximations used to balance the need for accurate explanations with the potential impact on the model's predictive performance. Future work could explore methods to minimize this trade-off, perhaps by using adaptive discretization techniques or by focusing on explaining models or parts of models where a high degree of accuracy is maintained even with the logical encoding.

\section{Directions for Future Work and Conclusion}

\subsection{Directions for Future Work}

Several avenues exist for future research to build upon and extend the proposed method:

\begin{itemize}
    \item   \textbf{Exploring Different SAT Solving Techniques and Optimizations:} Investigating the effectiveness of various SAT solving algorithms, including CDCL with different heuristics, and exploring optimizations tailored to the specific types of SAT problems generated from encoding deep learning models for video.

    \item   \textbf{Extending the Model to Handle Different Types of Video Understanding Tasks:} Applying and adapting the proposed method to a wider range of video understanding tasks beyond action recognition, such as video captioning, anomaly detection, and event recognition. Each task might require specific considerations in terms of feature representation and explanation queries.

    \item   \textbf{Investigating Methods for Generating More Human-Interpretable Explanations from the SAT-Based Framework:} Developing techniques to translate the logical explanations obtained from the SAT solver into more human-readable formats, such as natural language descriptions, visual summaries, or high-level logical rules. This would make the explanations more accessible and useful to end-users.

    \item   \textbf{Potential for Empirical Evaluation in Specific Video Understanding Domains:} Conducting empirical evaluations of the proposed model on real-world video datasets to assess its effectiveness in providing meaningful explanations and to evaluate its scalability for practical applications. This would involve selecting appropriate deep learning models for video understanding, designing relevant explanation queries, and analyzing the quality and computational cost of the generated explanations.

    \item   \textbf{Exploring Alternative Logical Frameworks:} Investigating the use of more expressive logical frameworks like SMT or first-order logic to encode deep learning models for video, potentially allowing for a more direct representation of continuous values and complex relationships. This would involve addressing the challenges associated with solving problems in these more expressive logics.
\end{itemize}

\subsection{Conclusion}

This paper has presented a novel formal SAT-based explanation model for deep learning in video understanding. The proposed method aims to address the limitations of existing XAI techniques by integrating the rigorous reasoning capabilities of SAT solvers with the principles of formal explainable AI. By formally encoding deep learning models and video data into a logical framework and formulating explanation queries as satisfiability problems, the method offers the potential to generate logic-based explanations with formal guarantees. The paper detailed the conceptual framework, the process of encoding deep learning models and video data, the formulation of "Why?" and "Why not?" questions, and a novel architecture integrating a SAT solver with a deep learning video understanding model. While challenges related to computational complexity and the representational power of propositional logic remain, the proposed approach offers a promising direction for enhancing the explainability of deep learning in the complex and critical domain of video understanding. Future work will focus on addressing these challenges, extending the model to various video understanding tasks, and developing methods for generating more human-interpretable explanations, ultimately aiming to increase trust and accountability in AI-driven video analysis.

\bibliographystyle{plain}
\bibliography{main} 
\end{document}